\documentclass[aps,preprint,tightenlines,superscriptaddress,showpacs]{revtex4}

\usepackage{graphicx} 
\usepackage{dcolumn}  

\usepackage{epsfig}
\usepackage{refmerge}
\usepackage{color}

\def \vec #1{\mbox{{\boldmath $#1$}}}

\def \B {{\cal B}}

\def \GeV {{\rm GeV}}
\def \MeV {{\rm MeV}}
\def \keV {{\rm keV}}
\def \eV {{\rm eV}}

\def \simgt {\stackrel{>}{\sim}}

\begin{document}

\preprint{\vbox{ \hbox{   }
                 \hbox{   }
                 \hbox{   }
                 \hbox{KEK Preprint 2008-50}
                 \hbox{Belle Preprint 2009-4}
}}                 

\title{ \quad\\[0.5cm]  High-statistics study of neutral-pion pair production\\
in two-photon collisions }

\affiliation{Budker Institute of Nuclear Physics, Novosibirsk}
\affiliation{University of Cincinnati, Cincinnati, Ohio 45221}
\affiliation{T. Ko\'{s}ciuszko Cracow University of Technology, Krakow}
\affiliation{The Graduate University for Advanced Studies, Hayama}
\affiliation{Gyeongsang National University, Chinju}
\affiliation{Hanyang University, Seoul}
\affiliation{University of Hawaii, Honolulu, Hawaii 96822}
\affiliation{High Energy Accelerator Research Organization (KEK), Tsukuba}
\affiliation{Hiroshima Institute of Technology, Hiroshima}
\affiliation{Institute of High Energy Physics, Chinese Academy of Sciences, Beijing}
\affiliation{Institute of High Energy Physics, Protvino}
\affiliation{Institute for Theoretical and Experimental Physics, Moscow}
\affiliation{J. Stefan Institute, Ljubljana}
\affiliation{Kanagawa University, Yokohama}
\affiliation{Korea University, Seoul}
\affiliation{Kyungpook National University, Taegu}
\affiliation{\'Ecole Polytechnique F\'ed\'erale de Lausanne (EPFL), Lausanne}
\affiliation{Faculty of Mathematics and Physics, University of Ljubljana, Ljubljana}
\affiliation{University of Maribor, Maribor}
\affiliation{University of Melbourne, School of Physics, Victoria 3010}
\affiliation{Nagoya University, Nagoya}
\affiliation{Nara Women's University, Nara}
\affiliation{National Central University, Chung-li}
\affiliation{National United University, Miao Li}
\affiliation{Department of Physics, National Taiwan University, Taipei}
\affiliation{H. Niewodniczanski Institute of Nuclear Physics, Krakow}
\affiliation{Nippon Dental University, Niigata}
\affiliation{Niigata University, Niigata}
\affiliation{Novosibirsk State University, Novosibirsk}
\affiliation{Osaka City University, Osaka}
\affiliation{Panjab University, Chandigarh}
\affiliation{Saga University, Saga}
\affiliation{University of Science and Technology of China, Hefei}
\affiliation{Seoul National University, Seoul}
\affiliation{Sungkyunkwan University, Suwon}
\affiliation{University of Sydney, Sydney, New South Wales}
\affiliation{Tata Institute of Fundamental Research, Mumbai}
\affiliation{Toho University, Funabashi}
\affiliation{Tohoku Gakuin University, Tagajo}
\affiliation{Department of Physics, University of Tokyo, Tokyo}
\affiliation{Tokyo Metropolitan University, Tokyo}
\affiliation{Tokyo University of Agriculture and Technology, Tokyo}
\affiliation{IPNAS, Virginia Polytechnic Institute and State University, Blacksburg, Virginia 24061}
\affiliation{Yonsei University, Seoul}
\author{S.~Uehara}\affiliation{High Energy Accelerator Research Organization (KEK), Tsukuba} 
\author{Y.~Watanabe}\affiliation{Kanagawa University, Yokohama} 
\author{H.~Nakazawa}\affiliation{National Central University, Chung-li} 
\author{H.~Aihara}\affiliation{Department of Physics, University of Tokyo, Tokyo} 
\author{K.~Arinstein}\affiliation{Budker Institute of Nuclear Physics, Novosibirsk}\affiliation{Novosibirsk State University, Novosibirsk} 
\author{V.~Aulchenko}\affiliation{Budker Institute of Nuclear Physics, Novosibirsk}\affiliation{Novosibirsk State University, Novosibirsk} 
\author{A.~M.~Bakich}\affiliation{University of Sydney, Sydney, New South Wales} 
\author{E.~Barberio}\affiliation{University of Melbourne, School of Physics, Victoria 3010} 
\author{K.~Belous}\affiliation{Institute of High Energy Physics, Protvino} 
\author{V.~Bhardwaj}\affiliation{Panjab University, Chandigarh} 
% ...

\collaboration{Belle Collaboration) \\
(To be published in Phys. Rev. D}

\begin{abstract}
The differential cross sections for the process 
$\gamma \gamma \to \pi^0 \pi^0$
have been measured in the kinematic range
0.6~GeV $< W < 4.1$~GeV, $|\cos \theta^*|<0.8$
in energy and pion scattering angle, respectively,
in the $\gamma\gamma$ center-of-mass system.
The results are based on a 223~fb$^{-1}$ data sample
collected with the Belle detector at the KEKB $e^+ e^-$ collider.
The differential cross sections are fitted in the energy region  
$1.7~\GeV < W <  2.5~\GeV$ to confirm the two-photon production 
of two pions in the G wave.
In the higher energy region,
we observe production of the  $\chi_{c0}$ charmonium state and obtain the
product of its two-photon decay width and branching fraction to 
$\pi^0\pi^0$.
We also compare the observed
angular dependence and ratios of cross sections for neutral-pion 
and charged-pion pair production to QCD models.
The energy and angular dependence above 3.1~GeV are compatible 
with those measured in the $\pi^+\pi^-$ channel, and in addition we find
that the cross section ratio,
$\sigma(\pi^0\pi^0)/\sigma(\pi^+\pi^-)$, is $0.32 \pm 0.03 \pm 0.05$
on average in the 3.1-4.1~GeV region. 
\end{abstract}

\pacs{13.20.Gd, 13.60.Le, 13.66.Bc, 14.40.Cs,14.40.Gx}
\maketitle
\tighten
\normalsize
{\renewcommand{\thefootnote}{\fnsymbol{footnote}}
\setcounter{footnote}{0}

\section{Introduction}
Measurements of exclusive hadronic final states in two-photon
collisions provide valuable information concerning the physics of light and 
heavy-quark resonances, perturbative and nonperturbative QCD, 
and hadron-production mechanisms.
So far, we have measured
the production cross sections for charged-pion 
pairs~\cite{mori,nkzw}, 
charged- and neutral-kaon pairs~\cite{abe,nkzw,wtchen}, 
and proton-antiproton pairs~\cite{kuo}.
We have also analyzed $D$-meson-pair production and observe a new
charmonium state~\cite{z3930}.
Recently, we have presented a measurement of neutral-pion pair
production based on a data sample corresponding to
an integrated luminosity of 95~fb$^{-1}$~\cite{pi0pi0}.
We have carried out an analysis in the energy range $W < 1.6~\GeV$
to extract information
on light quark resonances from the energy and angular
dependence of the differential cross sections, by
fitting to the resonance parameters of the $f_0(980)$, $f_2(1270)$ 
and additional hypothetical resonances.
The statistics of these measurements is 2 to 3 orders of
magnitude higher than in the pre-$B$-factory measurements~\cite{past_exp}, 
opening a new era in studies of two-photon physics.

Here we present measurements of the differential cross sections, 
$d\sigma/d|\cos \theta^*|$, for the process $\gamma \gamma \to \pi^0 \pi^0$
in a wide two-photon center-of-mass (c.m.) 
energy ($W$) range from 0.6 to 4.1~GeV,
and in the c.m. angular range, $|\cos \theta^*| <0.8$.
We use a 223~fb$^{-1}$ data sample, which is more than twice 
as large as that in our previous analysis~\cite{pi0pi0}.
We focus on the range $W > 1.4~\GeV$,
where the previous data was statistically limited.

In the intermediate energy range ($1.0~\GeV < W < 2.4~\GeV$), 
production of two pions is dominated by intermediate resonances.
For ordinary $q\bar{q}$ mesons in isospin conserving decays
to $\pi\pi$, the only allowed  $I^GJ^{PC}$ states
produced by two photons are  
$0^+$(even)$^{++}$, that is, $f_{J={\rm even}}$ mesons.
Several mesons with these quantum numbers are suggested by results of 
hadron-beam or charmonium decay experiments in the
1.5 - 2.2~GeV region.
However, none of them have been firmly established in two-photon processes,
which are sensitive to the internal quark structure of the meson. 
In addition, the $\pi^0\pi^0$ channel has two advantages in the study of
resonances: a smaller contribution from the 
continuum is expected in it than in the $\pi^+\pi^-$ channel;
and the angular coverage is larger ($|\cos \theta^*| < 0.8$ 
instead of 0.6).

At higher energies ($W > 2.4~\GeV$), we can invoke a quark model.
In leading-order calculations~\cite{bl,bc,chern}, which take into account
the spin correlation between quarks, the $\pi^0\pi^0$
cross section is predicted to be much smaller
than that of $\pi^+\pi^-$, suggesting a ratio 
of $\pi^0\pi^0$ to $\pi^+\pi^-$ cross sections around 0.04-0.07.
However, higher-order or
nonperturbative QCD effects can modify this prediction. 
For example, the handbag model, which considers soft hadron exchange,
predicts the same amplitude for the two processes,
and thus the expected ratio is 0.5~\cite{handbag}. 
Analyses of energy and angular distributions of the cross sections
are essential for determining the properties of the observed
resonances and for testing the validity of QCD models.

The organization of this article is as follows.
In Sec.~\ref{sec:appar}, a brief description of the Belle
detector is given.
Section~\ref{sec:cross} explains the procedure used to obtain differential
cross sections.
Section~\ref{sec:reson} is devoted to results on
the two-photon production of two pions in the G wave obtained
by fitting differential cross sections in the range $1.7~\GeV < W < 2.5~\GeV$.
Section~\ref{sec:highe} describes analyses at higher energy.
The topics included there are the angular dependence as a function of $W$, 
the observation of the $\chi_{c0}$ and $\chi_{c2}$ charmonia states
and the ratio of cross sections for $\pi^0 \pi^0$ to $\pi^+ \pi^-$
production.
Finally, Sec.~\ref{sec:concl} summarizes the results and 
presents the conclusion of this paper.

\section{Experimental apparatus}
\label{sec:appar}
We use a 223~fb$^{-1}$ data sample from the Belle experiment~\cite{belle}
at the KEKB asymmetric-energy $e^+e^-$ collider~\cite{kekb}.
The data were recorded at several $e^+e^-$ c.m. energies 
summarized in Table~\ref{tab:lum_data}.
The difference of the luminosity functions
(two-photon flux per $e^+e^-$-beam luminosity)
in the measured $W$ regions due to the difference of
the beam energies is small (maximum $\pm$ 4\%). 
We combine the results from the different beam energies. 
The effect on the cross section is less than 0.5\%.
\begin{center}
\begin{table}
\caption{Data sample: luminosities and energies}
\label{tab:lum_data}
\begin{tabular}{ccl} \hline \hline
~~$e^+e^-$ c.m. energy~~ & ~~~~Luminosity~~~~ & Comment \\
(GeV)& (fb$^{-1}$) & \\ \hline
10.58 & 179 & $\Upsilon(4S)$ runs \\ 
10.52 & 19 & continuum runs\\
10.36 & 2.9 & $\Upsilon(3S)$ runs \\
10.30 & 0.3 & continuum runs\\
10.86 & 21.7 & $\Upsilon(5S)$ runs \\ \hline
total & 223 & \\
\hline\hline
\end{tabular}
\end{table}
\end{center}

The analysis is carried out in the ``zero-tag'' mode, where
neither the recoil electron nor positron are detected. 
We restrict the virtuality of the incident photons to be small
by imposing strict transverse-momentum balance with respect to 
the beam axis for the final-state hadronic system.

A comprehensive description of the Belle detector is
given elsewhere~\cite{belle}. 
We mention here only those
detector components that are essential for the present measurement.
Charged tracks are reconstructed from hit information in 
the silicon vertex detector and the central drift chamber
 located in a uniform 1.5~T solenoidal magnetic field.
The detector solenoid is oriented along the $z$ axis, which points
in the direction opposite to that of the positron beam. 
Photon detection and
energy measurements are performed with a CsI(Tl) electromagnetic
calorimeter (ECL).

For this all-neutral final state, we require that there be no
reconstructed tracks coming from the vicinity of
the nominal collision point. 
Therefore, the central drift chamber is used for vetoing events with charged track(s). 
The photons from decays of two neutral pions are detected and their 
momentum vectors are measured by the ECL. 
The ECL is also used to trigger signal events.

\section{Deriving differential cross sections}
\label{sec:cross}
The event triggers, data processing, 
and event selection are the same as those described in Ref.~\cite{pi0pi0}.
We derive the c.m. energy $W$ of the two-photon collision 
from the invariant mass of the two-neutral-pion system.
We calculate the cosine, $| \cos \theta^*|$ of the $\pi^0$ scattering angle
in the $\gamma \gamma$ c.m. frame
for each event, using the $e^+e^-$ collision 
axis in the $e^+e^-$ c.m. frame as the reference axis for the polar angle.
The possible bias due to the unknown $\gamma \gamma$ collision axis is 
negligible.

\subsection{Data reduction}
We find that the signal candidates in the low energy
region ($W<1.2~\GeV$) are considerably contaminated
by background.
In order to separate the signal and background components, we study 
the $p_t$-balance 
distribution, i.e., the event distribution in $|\sum \vec{p}_t^*|$.
We estimate the $p_t$-unbalanced background component for $W<1.2~\GeV$ 
in the same manner as in the previous analysis~\cite{pi0pi0}
and subtract the yield in the signal region.
However, above 1.2~GeV, we cannot quantitatively determine the background
contamination because of the small background rate and low statistics of the
sample, as well as the uncertainty in the functional form 
for the signal shape.

Using the ratio of yields between the $p_t$-balanced and 
unbalanced regions, we can estimate the backgrounds. 
In Fig.~\ref{fig:fig1},
we plot the $W$ dependence of $R$ defined as:
\begin{equation}
R= \frac{Y(0.15~\GeV/c < |\sum \vec{p}_t^*| < 0.20~\GeV/c)}
  {Y (|\sum \vec{p}_t^*|<0.05~ \GeV/c)} \; ,
\label{eqn:r_pt}
\end{equation}
where $Y$ is the yield in the indicated $|\sum \vec{p}_t^*|$ region.
We integrate over all angles in this figure.
The main part of the $W$ dependence of $R$ comes from
the energy dependence of the momentum resolution. 
The expected ratio from the pure signal component 
is shown by the solid line.
The signal events for $e^+e^- \to e^+e^- \pi^0\pi^0$ are
generated using the TREPS code~\cite{treps}.
All Monte Carlo (MC) events are put through
the trigger and detector simulators and the event selection program.
The MC events are corrected for MC/data difference in
the $p_t$ resolution discussed in the next section. 
The excess of $R$ over the line ($\Delta R$) is expected to correspond
to the contribution from the $p_t$-unbalanced background.
The excess is relatively small above 1.0~GeV, although some fine structure
is visible there.
In the range 1.2--3.3~GeV, $\Delta R$ ranges between 0.00 and 0.08, and
above $3.3~\GeV$ it is in the range from 0.08 to 0.2.
From the $R$ values, we estimate that the background 
contamination in the signal region is $\sim R/4$, which
is  smaller than 3\% for 1.5 - 3.3~GeV and around 3\% for 3.6 - 4.1~GeV. 
We subtract 3\% for 3.6 - 4.1~GeV, and assign a 3\% systematic 
error from this source for the full 1.5 - 4.1~GeV range. 

We estimate the invariant-mass resolution from studies 
of signal-MC and experimental distribution. 
The true $W$ distribution in the range 0.9~GeV $<W<$ 2.4 ~GeV is 
obtained by unfolding the differential cross sections as described in 
Ref.~\cite{pi0pi0}.
For lower energies, $W < 0.9~\GeV$, the effect of the migration is 
expected to be small because the invariant-mass resolution is
much smaller than the bin width. 
For higher energies, $W > 2.4~\GeV$, 
where the statistics is relatively low and unfolding would enlarge 
the errors, we adopt a rather wide bin width (100 MeV) without unfolding.
A total of $2.90 \times 10^6$ events are selected 
in the region of $0.6~\GeV < W < 4.1~\GeV$
and $|\cos \theta^*| < 0.8$.

\begin{figure}
\centering
\includegraphics[width=9cm]{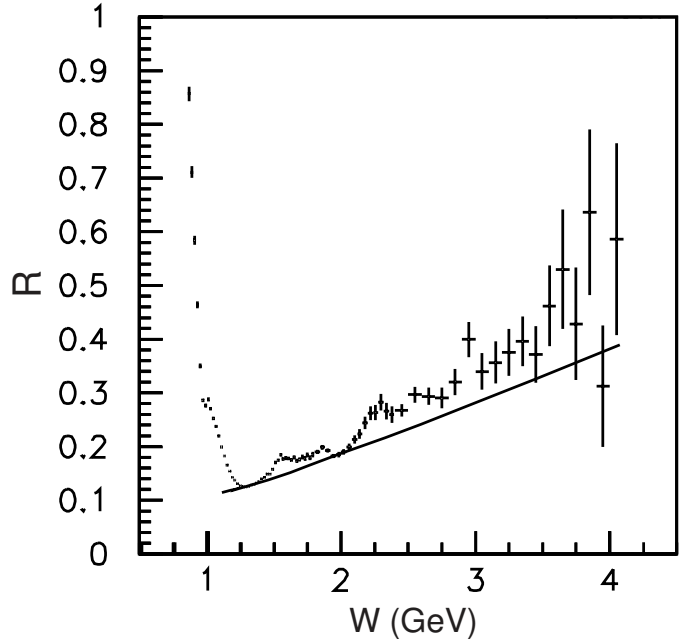}
\centering
\caption{
The yield ratio $R$ in the $p_t$-unbalanced
bin to the $p_t$-balanced (signal) bin (see text for the
exact definition) for the experimental data.
The solid line shows the signal component obtained
from the signal MC and corrected taking into account the poorer momentum
resolution in experimental data.}
\label{fig:fig1}
\end{figure}

\subsection{Calculation of differential cross sections}
We determine the efficiency for the signal using a full MC simulation. 
The MC signal events generated using the TREPS code~\cite{treps} are 
isotropically distributed in $|\cos \theta^*|$ at 58 fixed $W$ points 
between 0.5 and 4.5~GeV.
The angular distribution at 
the generator level does not play a role in the efficiency
determination, because we calculate the efficiencies separately 
in each $|\cos \theta^*|$ bin with a 0.05 width.
Samples of $4 \times 10^5$ events are generated at each $W$ point. 
Two sets of different background conditions, which were extracted from
the beam collision data are embedded in the signal-MC
data in the detector simulation. 
To minimize statistical fluctuations in the MC calculation, 
we fit the numerical results of the trigger efficiency to a
two-dimensional empirical function in $(W, |\cos \theta^*|)$.

The efficiency calculated from the signal-MC events is 
corrected for a systematic difference of the peak 
widths in the $p_t$-balance distributions found between
the experimental data and the MC events, which is attributed
to a difference in the momentum resolution for $\pi^0$'s.
The correction factor is typically 0.95.

The differential cross section for each
($W$, $|\cos \theta^*|$) point is given by:
\begin{equation}
\frac{d\sigma}{d|\cos \theta^*|} =
\frac{\Delta Y - \Delta B}{\Delta W \Delta |\cos \theta^*| 
\int{\cal L}dt L_{\gamma\gamma}(W)  \eta } \; ,
\label{eqn:diffc}
\end{equation}
where $\Delta Y$ and $\Delta B$ are the signal yield and
the estimated $p_t$-unbalanced background in the bin, 
$\Delta W$ and $\Delta |\cos \theta^*|$ are the bin widths, 
$\int{\cal L}dt$ and  $L_{\gamma\gamma}(W)$ are
the integrated luminosity and two-photon luminosity function
calculated by TREPS~\cite{treps}, respectively, and  $\eta$ is the efficiency 
including the correction described above.
The bin sizes for $W$ and $\Delta |\cos \theta^*|$ are summarized in
Table~\ref{tab:binsize}.
\begin{center}
\begin{table}
\caption{Bin sizes}
\label{tab:binsize}
\begin{tabular}{lcc} \hline \hline
$W$ range & $\Delta W$ & $\Delta |\cos \theta^*|$ \\ 
(GeV) & (GeV) & \\\hline
0.6 -- 1.8 & 0.02 & 0.05 \\
1.8 -- 2.4 & 0.04 & 0.05 \\
2.4 -- 4.1 & 0.10 & 0.05 \\
\hline\hline
\end{tabular}
\end{table}
\end{center}

Figure~\ref{fig:fig2} 
shows the $W$ dependence of the
cross section integrated over $|\cos \theta^*| < 0.8$. 
We have removed the bins in the range $3.3~\GeV < W < 3.6~\GeV$, 
because we cannot separate the $\chi_{c0}$ and $\chi_{c2}$  
components and the continuum in a model-independent way 
due to the finite mass resolution and insufficient statistics of 
the measurement. 
The cross section in this region is 
discussed in detail in Sec.~\ref{sec:highe}.

We show the angular dependence of the differential cross sections
at several $W$ points in Fig.~\ref{fig:fig3}.
Note that the cross sections in neighboring bins after the unfolding are 
no longer independent of each other
in either central values or size of errors.

\begin{figure}
\centering
\includegraphics[width=9cm]{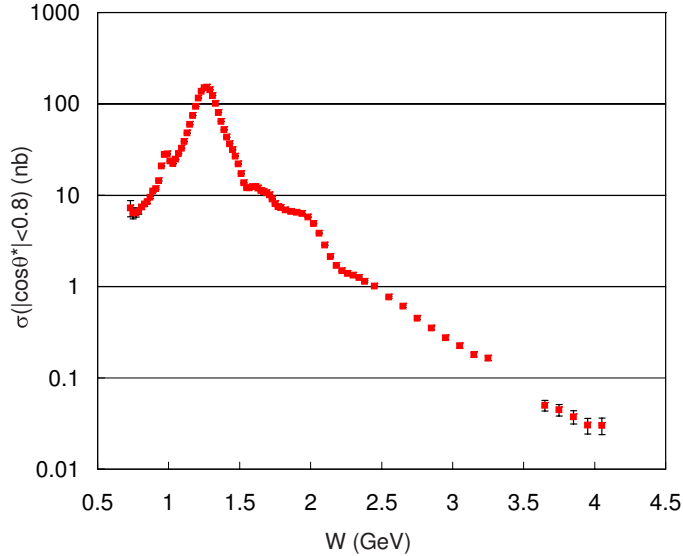}
\centering
\caption{The integrated cross section
in the angular regions $|\cos \theta^*|<0.8$.
Data points in bins near 3.5~GeV are not shown because of uncertainty
from the $\chi_{cJ}$ subtraction.}
\label{fig:fig2}
\end{figure}

\begin{figure}
\centering
\includegraphics[width=14cm]{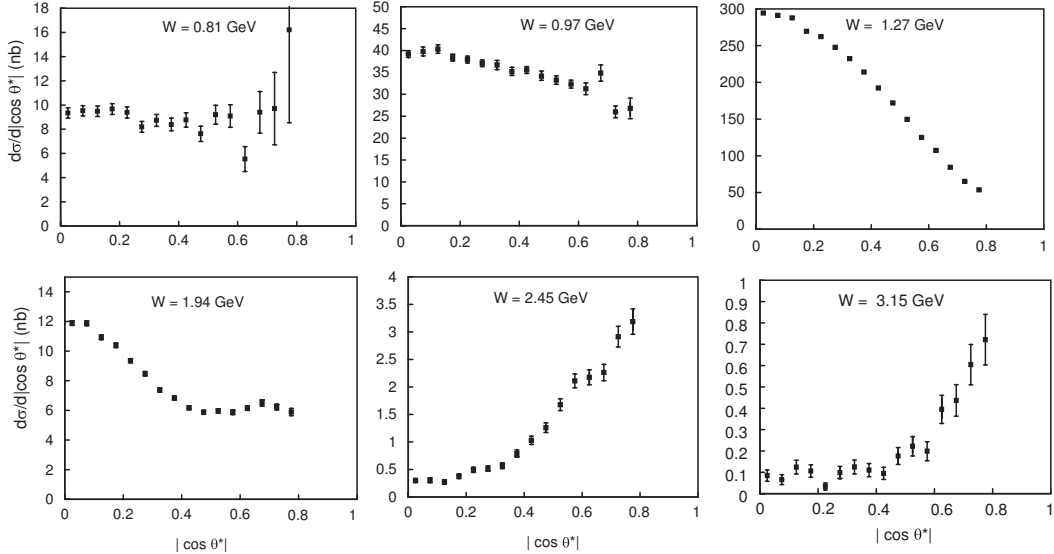}
\centering
\caption{
The differential cross sections for $W$ points indicated.
The bin sizes are summarized in Table~\ref{tab:binsize}.}
\label{fig:fig3}
\end{figure}

The systematic uncertainties for the cross sections
arise from various sources; they
are listed in Table~\ref{tab:sysdcs} together with the estimated values.
Uncertainties from the unfolding procedure, using the single value
decomposition approach in Ref.~\cite{svdunf},
are estimated by varying the effective-rank parameter
of the decomposition within reasonable bounds.

The total systematic error is obtained by adding the uncertainties
in quadrature and is about 10\% in the intermediate $W$ region
(1.04~GeV $<W< 3.0$~GeV). 
It becomes much larger at lower $W$.
At higher $W$, the systematic error is rather
stable, typically about 11\%.
\begin{center}
\begin{table}
\caption{
Systematic errors for the differential cross sections.
Ranges of errors are shown when they depend on $W$.}
\label{tab:sysdcs}
\begin{tabular}{lc} \hline \hline
Source & Error (\%) \\ \hline
Trigger efficiency & 4 -- 30\\
$\pi^0$ reconstruction efficiency & 6 \\
$p_t$-balance cut  & 1.5 -- 5 \\
Background subtraction & 0 -- 40 \\
Luminosity function & 4 -- 5 \\
Overlapping hits from beam background & 2 -- 4 \\ 
Other efficiency errors & 4 \\
Unfolding procedure & 0 -- 4 \\ \hline
Overall & typ. 10 -- 11 \\
\hline\hline
\end{tabular}
\end{table}
\end{center}
}

\section{Study of G-wave activity}
\label{sec:reson}
Previously, we have obtained a reasonable fit to a simple model
of resonances and smooth backgrounds in the energy region
$0.8~\GeV < W < 1.6~\GeV$ from the differential cross sections
of $\gamma \gamma \rightarrow \pi^0 \pi^0$ with a 95~fb$^{-1}$
data sample~\cite{pi0pi0}.
The clear $f_0(980)$ peak and the large contribution from
the $f_2(1270)$ can be fitted with 
parameters determined from $\pi^+ \pi^-$ data~\cite{mori}.

In this section, we concentrate on the 
G wave, in particular, the $f_4(2050)$ resonance, 
whose existence is well established, but whose production
in two-photon collisions has never been positively identified.
We fit the energy region $1.7~\GeV < W < 2.5~\GeV$ 
using a high-statistics sample of 223~fb$^{-1}$ that contains
2.3 times more events than in the previous experiment~\cite{pi0pi0};
the number of events in this region is 155k.

\subsection{Parametrization of Partial Wave Amplitudes}
\begin{figure}
 \centering
\includegraphics[width=7.5cm]{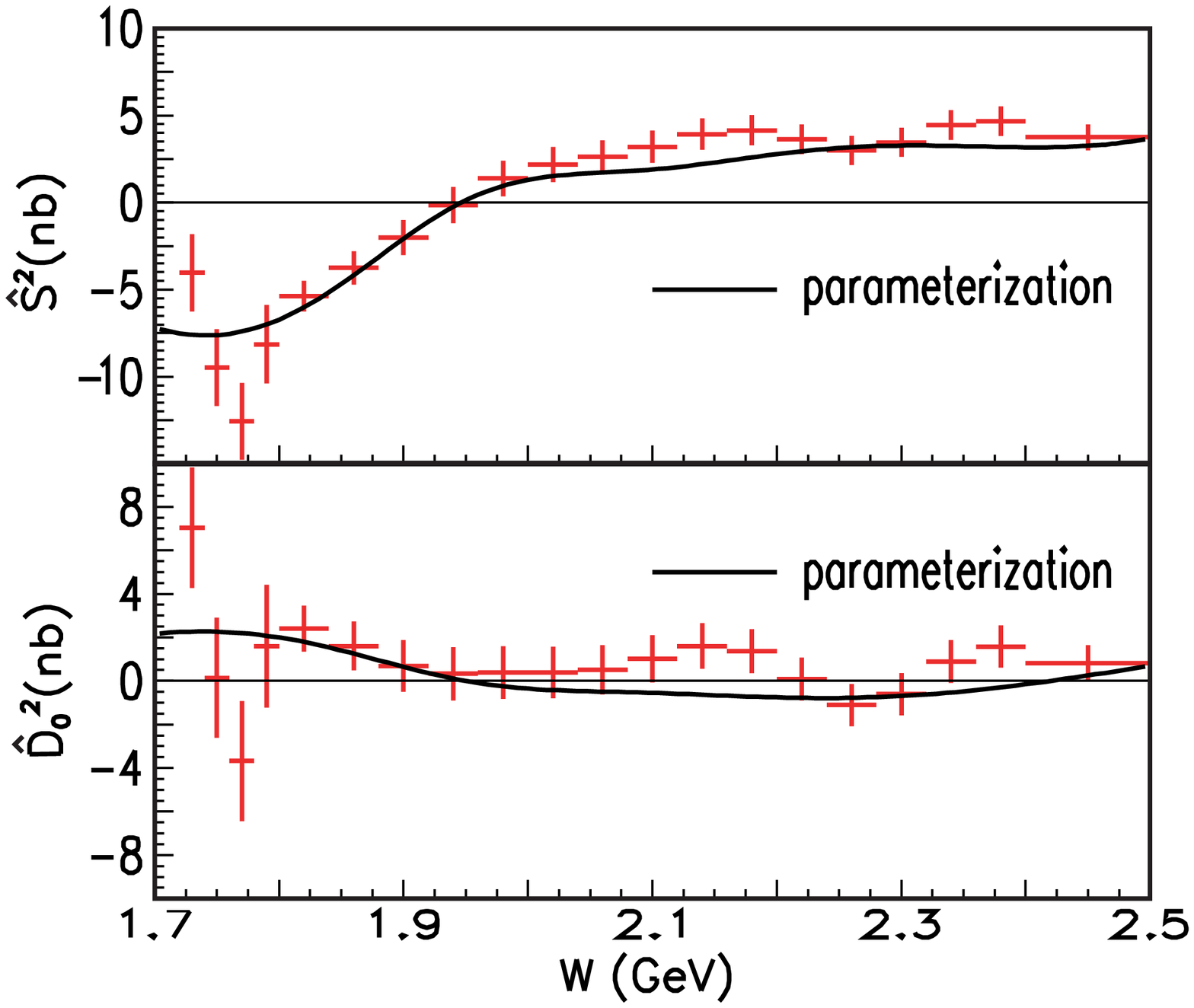}
\includegraphics[width=7.5cm]{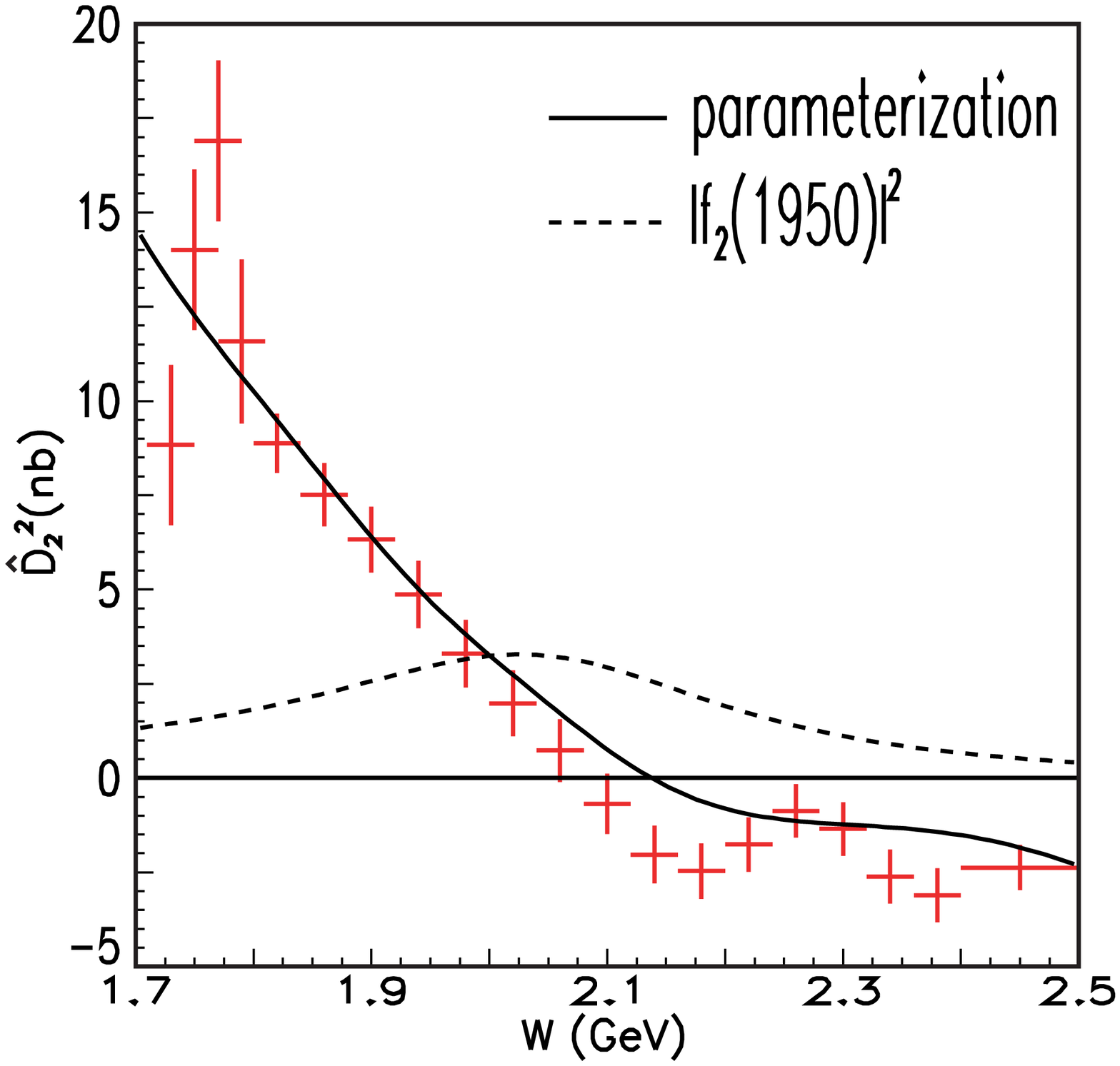}
 \caption{Spectrum of $\hat{S}^2$ (top section of left-hand panel), 
$\hat{D_0}^2$ (bottom section of left-hand panel)
and  $\hat{D_2}^2$ (right panel)  for $1.7~\GeV < W < 2.5~\GeV$ 
and results of parametrization (see text).
The error bars shown are diagonal statistical errors.}
\label{fig:fig4}
\end{figure}

\begin{figure}
 \centering
\includegraphics[width=7.5cm]{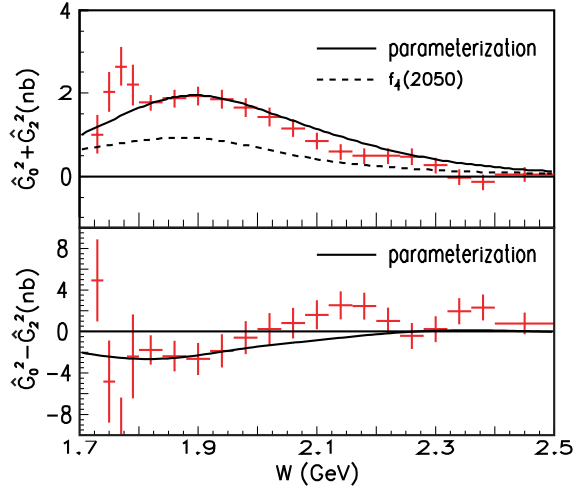}
 \caption{Spectrum of $\hat{G_0}^2 + \hat{G_2}^2$ (top section) 
and $\hat{G_0}^2 - \hat{G_2}^2$ (bottom section)
 for $1.7~\GeV < W < 2.5~\GeV$ and fitted curves (see text).
The error bars shown are diagonal statistical errors.}
\label{fig:fig5}
\end{figure}

In the energy region $W \leq 3$~GeV, $J > 4$ partial waves (the next is 
$J=6$) may be neglected so that only S, D and G waves are to be considered.
The differential cross section can be expressed as:
\begin{equation}
\frac{d \sigma}{d \Omega} (\gamma \gamma \to \pi^0 \pi^0)
 = \left| S \: Y^0_0 + D_0 \: Y^0_2  + G_0 \: Y^0_4 \right|^2 
+ \left| D_2 \: Y^2_2  + G_2 \: Y^2_4 \right|^2 \; ,
\label{eqn:diff}
\end{equation}
where $D_0$ and $G_0$ ($D_2$ and $G_2$) denote the helicity 0 (2) components
of the D and G waves, respectively, and $Y^m_J$ are the spherical harmonics.
Since the $|Y^m_J|$s are not independent,
partial waves cannot be separated using measurements of
differential cross sections alone.
To overcome this problem, we write Eq.~(\ref{eqn:diff}) as
\begin{equation}
\frac{d \sigma}{4 \pi d |\cos \theta^*|} (\gamma \gamma \to \pi^0 \pi^0)
 = \hat{S}^2 \: |Y^0_0|^2  + \hat{D}_0^2 \: |Y^0_2|^2
+ \hat{D}_2^2  \: |Y^2_2|^2 \, 
+ \hat{G}_0^2  \: |Y^0_4|^2  \, 
+ \hat{G}_2^2  \: |Y^2_4|^2  \, .
\label{eqn:diff2}
\end{equation}
The amplitudes $\hat{S}^2$, etc. correspond to the cases where 
interference terms are neglected; they can be expressed in terms of 
$S,~D_0,~D_2,~G_0$, and $G_2$~\cite{pi0pi0}. 
Since squares of spherical harmonics 
are independent of one another, we can fit differential cross 
sections at each $W$ to obtain 
$\hat{S}^2,~\hat{D}^2_0,~\hat{D}^2_2,~\hat{G}^2_0$, and $\hat{G}^2_2$.
For $|\cos{\theta^*}|<0.7$, the $|Y^0_4|^2$ and $|Y^2_4|^2$ terms
are nearly equal, 
so we fit $\hat{G}^2_0+\hat{G}^2_2$ and  $\hat{G}^2_0-\hat{G}^2_2$
instead. 
The resulting spectra are shown in Figs.~\ref{fig:fig4} and \ref{fig:fig5}.

We parametrize the partial wave amplitudes
in terms of resonances and smooth ``backgrounds''.
Once the functional forms of the amplitudes are fixed, we can use 
Eq.~(\ref{eqn:diff}) to fit the differential cross sections.
From Fig.~\ref{fig:fig5}, it appears that the 
G wave contributions are nonzero for $W \simgt 1.8~\GeV$ and are 
dominated by the G$_2$ wave.
Here we assume (and check the necessity of) including
the $f_4(2050)$ in the G$_2$ wave.
Since the G$_2$ wave interferes with the D$_2$ wave,
we include the resonance $f_2(1950)$,
which is known to couple to two photons~\cite{abe, pdg}.
There are several other resonances that might couple to $\gamma \gamma$
and $\pi \pi$ in this mass region, which are listed in Ref.~\cite{pdg}.
Here we assume that the $f_2(1950)$ is just an empirical 
parametrization representing these other resonances;
we denote it here as the ``$f_2(1950)$''.

We parametrize the partial waves as follows:
\begin{eqnarray}
S &=& B_S , \nonumber \\
D_0 &=& B_{D0}, \nonumber \\
D_2 &=& A_{\mbox{``}f_2(1950) \mbox{''}} e^{i \phi_{2}} + B_{D2}, \nonumber \\
G_0 &=& 0, \nonumber \\
G_2 &=& A_{f_4(2050)} e^{i \phi_{4}} + B_{G2} ,
\label{eqn:param2}
\end{eqnarray}
where $A_{\mbox{``}f_2(1950)\mbox{''}}$ and $A_{f_4(2050)}$ are the
amplitudes of the corresponding resonances;
$B_S$, $B_{D0}$, $B_{D2}$ and $B_{G2}$ are 
nonresonant (background) amplitudes for 
S, D$_0$, D$_2$ and G$_2$ waves; 
and $\phi_{2}$ and $\phi_{4}$
are the phases of resonances relative to background amplitudes.
We assume that $G_0 = 0$ and that $G_2$ consists only of the $f_4(2050)$
and a smooth background.

The relativistic Breit-Wigner resonance amplitude
$A_R(W)$ for a spin-$J$ resonance $R$ of mass $m_R$ is given by
\begin{eqnarray}
A_R^J(W) &=& \sqrt{\frac{8 \pi (2J+1) m_R}{W}} 
\frac{\sqrt{ \Gamma_{\rm tot} \Gamma_{\gamma \gamma} \B(\pi^0 \pi^0)}}
{m_R^2 - W^2 - i m_R \Gamma_{\rm tot}} \; .
\label{eqn:arj}
\end{eqnarray}
The resonance parameters given in Ref.~\cite{pdg} for the 
$f_2(1950)$ and $f_4(2050)$
are summarized in Table~\ref{tab:param}.
We assume an energy-independent width for the ``$f_2(1950)$'' and $f_4(2050)$ 
because most of their individual decay fractions are unknown.

\begin{center}
\begin{table}
\caption{Parameters of the $f_2(1950)$ and 
$f_4(2050)$~\cite{pdg}.}
\label{tab:param}
\begin{tabular}{lcccc} \hline \hline
Parameter  & $f_2(1950)$  & $f_4(2050)$ & Unit \\
\hline
Mass & $1944 \pm 12$ & $2018 \pm 11$ & MeV/$c^2$ \\
Width & $472 \pm 18$ & $237 \pm 18$ & MeV \\
${\cal B} (\pi \pi)$ & seen & $17.0 \pm 1.5$ & \% \\
${\cal B} (K \bar{K})$ & seen & $0.68^{+0.34}_{-0.18}$
& \% \\
${\cal B} (\eta \eta)$ & seen & $0.21 \pm 0.08$& \% \\
${\cal B} (\gamma \gamma)$  & seen & unknown &  \\
\hline\hline
\end{tabular}
\end{table}
\end{center}

The background amplitudes are parametrized as follows.
\begin{eqnarray}
B_S &=&  a_{sr} (W - W_0)^2  + b_{sr} (W - W_0) + c_{sr} 
+ i \left( a_{si} (W - W_0)^2  + b_{si} (W - W_0) + c_{si} \right),
\nonumber \\
B_{D0} &=& a_0 (W - W_0)^2 + b_0 (W - W_0) + c_0 , \nonumber \\
B_{D2} &=& a_{2r} (W - W_0)^2 + b_{2r} (W - W_0) + c_{2r} 
+ i\left( a_{2i} (W - W_0)^2  + b_{2i} (W - W_0) + c_{2i} \right), 
\nonumber \\
B_{G2} &=& a_{g} (W - W_0)^2  + b_{g} (W - W_0) + c_{g}
\label{eqn:para4}
\end{eqnarray}
where $W_0 = 1.7$~GeV.
The background amplitudes $D_0$ and $G_2$
are taken to be real by definition.
The other background amplitudes are assumed to be quadratic in $W$ for both
their real and imaginary parts.
We fix $B_{G2}=0$ at $W=1.7~\GeV$
($c_{g} = 0$) to reduce the number
of parameters; leaving $c_{g}$ free does not improve the fits.

\subsection{Fit results}
We minimize $\chi^2$ defined as
\begin{equation}
\chi^2 = \sum_{i,j} \left( \frac{ \frac{d \sigma}{d |\cos \theta^*|}
(W_i,|\cos \theta^*|_j)_{\rm data}
- \frac{d \sigma}{d |\cos \theta^*|} (W_i,|\cos \theta^*|_j)_{\rm pred.} }
{ \Delta \frac{d \sigma}{d |\cos \theta^*|}
(W_i,|\cos \theta^*|_j)_{\rm data} } \right)^2 \; ,
\label{eqn:chisq}
\end{equation}
where the summation is over $(W_i, |\cos \theta^*|_j)$ bins, 
$d \sigma/d |\cos \theta^*|(W_i,|\cos \theta^*|_j)_{{\rm data} \; 
{\rm (pred.)}}$
is the cross section data (prediction using Eq.(\ref{eqn:param2})) at
a bin $(W_i,|\cos \theta^*|_j)$,
and the denominator is the estimated statistical error.

When the mass and width of the $f_4(2050)$
are fixed to the values given in the
PDG tables~\cite{pdg} as summarized in Table~\ref{tab:param}, then
the fit is very poor yielding $\chi^2 \; (ndf) = 594.4 \; (313)$
(see Table~\ref{tab:fit1}).
This is to be compared with $323.2 \; (311)$ obtained when the
mass and width are floated.
In this paper we quote the results of the fits with the mass and width 
of the $f_4(2050)$ as free parameters.

\begin{figure}
 \centering
\includegraphics[width=12cm]{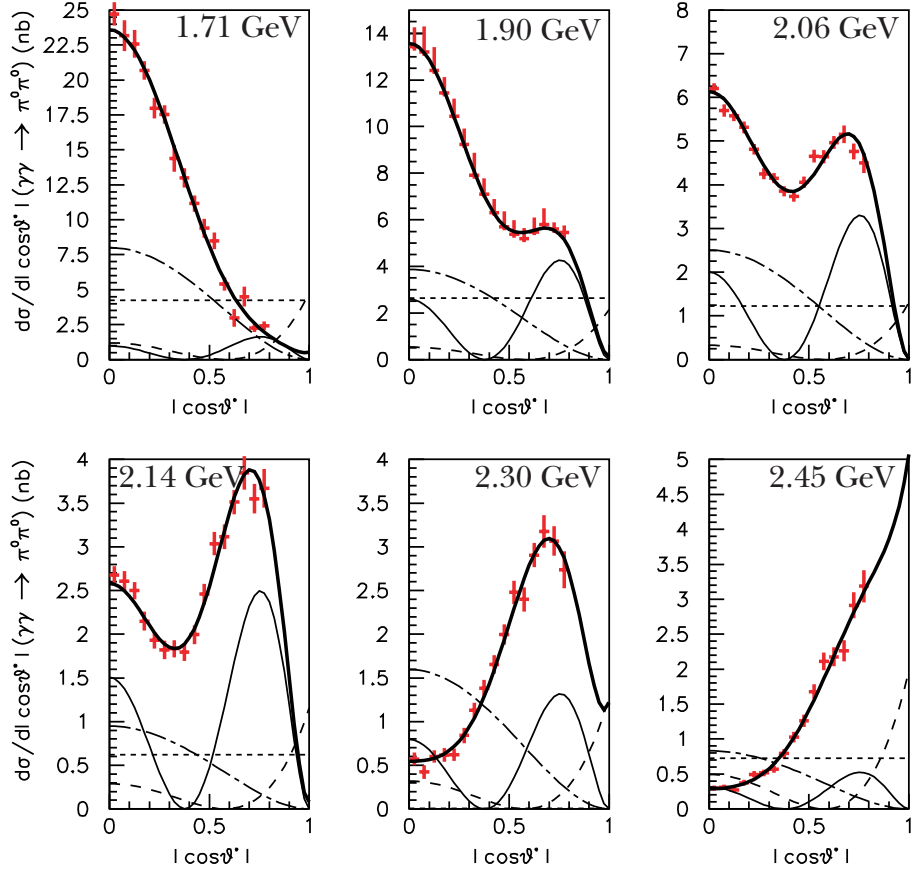}
 \caption{Measurements of $d \sigma / d |\cos \theta^*|$ (nb)
(data points) and results of the fit (thick solid line) 
for the $W$-bins indicated. 
The dotted, dashed, dot-dashed and thin lines indicate
$|S|^2$, $4 \pi |D_0 Y_2^0|^2$, $4 \pi |D_2 Y_2^2|^2$, $4 \pi |G_2 Y_4^2|^2$, 
respectively.}
\label{fig:fig6}
\end{figure}

\begin{figure}
 \centering
\includegraphics[width=8cm]{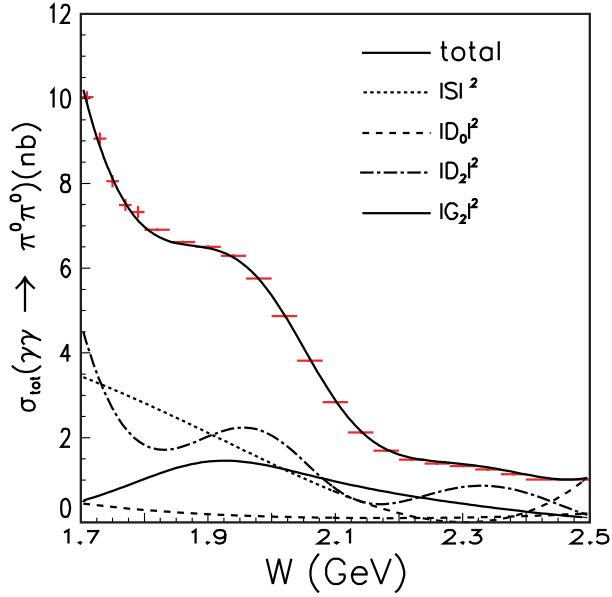}
 \caption{Total cross section ($|\cos \theta^*| < 0.8$) (nb)) and 
the results of the nominal fit (curves).}
\label{fig:fig7}
\end{figure}

Here the unfolded differential cross sections are fitted.
Fits are performed 1000 times for each study with 
randomly-generated initial values for the parameters 
A unique solution with good quality ($\chi^2/ndf = 1.04$)
is repeatedly found (``nominal fit'').
The fit results are shown in Fig.~\ref{fig:fig6} 
for the differential cross sections, in Fig.~\ref{fig:fig7} 
for the total cross section
and in Figs.~\ref{fig:fig4} and \ref{fig:fig5}
for the $\hat{S}^2$, etc.
Since the two-photon coupling of the $f_4(2050)$ has not been measured
before, a fit without this resonance is also given in Table~\ref{tab:fit1}.
The fit quality is unacceptable, strongly indicating that the $f_4(2050)$ 
has a nonzero two-photon coupling.
A fit without the ``$f_2(1950)$'' is also made giving a much worse fit 
and is included in Table~\ref{tab:fit1}.

We have performed additional fits to investigate
whether we can conclude that the $f_4(2050)$
is mainly produced in 
the helicity-2 state.
Note that the angular dependence of $Y_4^0$ and $Y_4^2$ is very similar
for $|\cos \theta^*| < 0.7$ and hence it is expected to be rather difficult 
to distinguish G$_0$ and G$_2$ waves.
A fit where the role of G$_0$ and G$_2$ is interchanged
(i.e. by setting $G_2 = 0$ and by including the $f_4(2050)$ 
and background in $G_0$) yields $\chi^2 =448.2$, which can be
compared to 323.2 for the nominal fit.
However, more reasonable fits are obtained
when two more parameters are introduced in the $G_2$ background
(a second-order polynomial real amplitudes or linear amplitudes for both 
real and imaginary parts,
which are set to zero at $W = 1.7~\GeV$).
Here the fits where the role of $G_2$ and $G_0$ have been interchanged 
are also compared.
In each case, the $f_4(2050)$ in a G$_2$ wave is favored over that in 
a G$_0$ wave with a $\chi^2$ differences of about 6 and 26, respectively.
Thus, the helicity-2 production of the $f_4(2050)$ is favored
but not conclusively.

According to Fig.~\ref{fig:fig7}, 
the $|D_2|^2$ term
has an enhancement around $W = 2.35~\GeV$,
which might be identified as the $f_2(2300)$.
To study this possibility, a fit is made including the $f_2(2300)$.
The conclusion is, however, that we have no sensitivity to it;
the fit does not improve significantly by its inclusion.
We believe that the enhancement arises from the ``$f_2(1950)$''
and its interference with the G$_2$ wave and underlying continuum, i.e.,
a fit without the ``$f_2(1950)$'' gives a smooth $D_2$ amplitude 
(with much worse $\chi^2$ (Table~\ref{tab:fit1})). 


\begin{center}
\begin{table}
\caption{Fitted parameters} 
\label{tab:fit1}
\begin{tabular}{lccccc} \hline \hline
Parameter & Nominal & Fixed $f_4(2050)$ & No $f_4(2050)$ 
& No \mbox{``}$f_2(1950)$\mbox{''} & Unit\\ \hline
Mass$(f_4(2050))$ &  $1885^{+14}_{-13}$ & 2025 (fixed)
& $\cdots$ & $2052 \pm 6$ & MeV/$c^2$ \\
$\Gamma_{\rm tot}(f_4(2050))$ & $453 \pm 20$ & 225 (fixed)
& $\cdots$ & $257^{+8}_{-7}$ & MeV\\
$\Gamma_{\gamma \gamma} {\cal B}(\pi^0 \pi^0)$ &  $7.7^{+1.2}_{-1.1}$ 
& $11.8 \pm 0.2$ & 0 (fixed) &  $14.2^{+0.9}_{-0.8}$ & eV\\
\hline
Mass$(\mbox{``}f_2(1950)\mbox{''})$ & $2038^{+13}_{-11}$  & $2026^{+2}_{-1}$  
& $2114^{+11}_{-13}$  & $\cdots$ & MeV/$c^2$\\
$\Gamma_{\rm tot}(\mbox{``}f_2(1950)\mbox{''})$ & $441^{+27}_{-25}$ 
& $237 \pm 4$ & $587^{+20}_{-1}$ & $\cdots$ & MeV\\
$\Gamma_{\gamma \gamma} {\cal B}(\pi^0 \pi^0)$
&  $54^{+23}_{-14}$  &  $76^{+48}_{-46}$ 
&  $334 ^{+79  }_{-77  }$ & 0(fixed) & eV \\
\hline
$\chi^2 \; (ndf)$ & 323.2 (311) & 594.4 (313) 
& 1397.8 (315) & 2306.8 (315) & \\
\hline\hline
\end{tabular}
\end{table}
\end{center}

\subsection{Study of systematic errors}
Various sources of systematic errors on the parameters are 
considered such as dependence on the fitted region, normalization errors 
of the differential cross sections, 
assumptions on the background amplitudes,
and uncertainties from the unfolding procedure.

For each study, a fit is made allowing all the parameters to vary.
The differences of the fitted parameters from the nominal values
are quoted as systematic errors.
Again, 1000 sets of randomly generated input parameters are prepared
for each study
and fitted to search for the true minimum and for possible multiple solutions.
Unique solutions are found repeatedly in all the cases.
Once the solutions are found,
several tens of repeated minimizations are needed to obtain fits that converge.
With many parameters ($24 - 26$ here) to be fitted, 
the approach to the minimum is rather slow.

Two fitting regions are tried: a higher one
($ 1.74~\GeV \leq W \leq 2.60~\GeV$)
and a lower one ($ 1.66~\GeV \leq W \leq 2.40~\GeV$).
The normalization error studies are divided into those from uncertainties of 
the overall normalization and those from distortion of the spectra in both
$|\cos \theta^*|$ and $W$.
For the overall normalization errors, fits are made with 
differential cross sections multiplied by 
$(1 \pm \sigma_{\epsilon(W, |\cos \theta^*|)})$, 
where $\sigma_{\epsilon}$ is the relative efficiency error.
For distortion studies, $\pm 4$\%  ($\pm 3$\%)
errors are assigned over the $|\cos \theta^*|$ ($W$) range
and differential cross sections are distorted by multiplying by 
$1 \pm 0.1 |\cos \theta^*| \mp 0.04$
 ($1 \pm 0.075 W \mp 0.1575$).

For studies of the background (BG) amplitudes, each of the waves 
is changed to a first- or a third-order polynomial except for the 
G$_0$ wave, where a first-order polynomial is introduced for both the
real and imaginary parts of the amplitude.
Parametrization uncertainties due to the phase convention where the D$_0$
and G$_2$ background amplitudes
are taken to be real are estimated by making the S and D$_2$ real
instead and by introducing imaginary parts for the D$_0$ and D$_2$ terms.

\begin{table}[h]
\caption{Systematic errors}
\label{tab:syser}
\begin{center}
\begin{tabular}{l|ccc|ccc} \hline \hline 
& \multicolumn{3}{c|}{$f_4(2050)$} 
& \multicolumn{3}{c}{\mbox{``}$f_2(1950)$\mbox{''}} 
\\ \cline{2-7}
Source & ~~Mass~~ & ~~~$\Gamma_{\rm tot}$~~~ 
& $\Gamma_{\gamma \gamma} {\cal B}_{\pi^0 \pi^0}$ 
 & Mass & $\Gamma_{\rm tot}$& $\Gamma_{\gamma \gamma} {\cal B}_{\pi^0 \pi^0}$\\
& (MeV/$c^2$) & ~~~(MeV)~~~ & ~~(eV)~~ &(MeV/$c^2$)& ~~~(MeV)~~~ & ~~(eV)~~ 
\\ \hline 
$W$-range & $^{+6}_{-0}$ & $^{+0}_{-28}$ & $^{+4.3}_{-0}$ 
& $^{+0}_{-14}$ & $^{+0}_{-87}$ & $^{+0}_{-27}$  \\
Normalization& $^{+0}_{-2}$ & $^{+0}_{-1}$ & $^{+1.0}_{-0.8}$ 
 & $^{+2}_{-0}$ & $^{+5}_{-0}$ & $^{+11}_{-3}$ \\
Bias:$|\cos \theta^*|$ & $^{+15}_{-16}$ & $^{+0}_{-0.7}$ 
& $^{+2.7}_{-2.4}$ & $^{+2}_{-0}$ & $^{+5}_{-3}$ & $^{+12}_{-3}$ \\
Bias:$W$& $^{0}_{-1}$ & $\pm 1$ & $\pm 0.2$ 
 & $^{+3}_{-2}$ & $^{+2}_{-0}$ & $^{+1}_{-0}$ \\
Unfolding & $^{+35}_{-0}$ & $^{+0}_{-68}$ & $^{+0}_{-3.0}$ 
& $^{+0}_{-44}$ & $^{+0}_{-84}$ & $^{+0}_{-36}$ 
\\  \hline 
BG: ${\rm Re} S$ & $^{+50}_{-0}$ & $^{+0}_{-72}$ & $^{+0}_{-3.1}$ 
 & $^{+0}_{-46}$ & $^{+9}_{-88}$ & $^{+0}_{-37}$ \\
BG: ${\rm Im} S$ & $^{+0}_{-1}$ & $^{+2}_{-7}$ & $^{+0}_{-0.1}$ 
 & $\pm 1$ & $^{+9}_{-0}$ & $^{+9}_{-0}$ \\
BG: $D_0$ & $^{+1}_{-13}$ & $^{+9}_{-15}$ & $^{+2.7}_{-0}$ 
 & $^{+3}_{-1}$ & $^{+3}_{-4}$ & $^{+22}_{-0}$ \\
BG: ${\rm Re} D_2$ & $^{+36}_{-0}$ & $^{+29}_{-2}$ & $^{+11.9}_{-0.6}$ 
 & $^{+0}_{-22}$ & $^{+24}_{-94}$ & $^{+16}_{-24}$ \\
BG: ${\rm Im} D_2$ & $^{+0}_{-12}$ & $^{+0}_{-12}$ & $^{+3.1}_{-0}$ 
 & $^{+0}_{-4}$ & $^{+0}_{-13}$ & $^{+11}_{-0}$ \\
BG: $G_0$ & $^{+20}_{-0}$ & $^{+0}_{-23}$ & $^{+0}_{-1.0}$ 
& $^{+2}_{-21}$ & $^{+6}_{-49}$ & $^{+1}_{-19}$ \\
BG: $G_2$ & $^{+205}_{-0}$ & $^{+7}_{-69}$ & $^{+19.1}_{-0}$ 
 & $\pm 10$ & $^{+0}_{-54}$ & $^{+377}_{-13}$ \\
BG: Real D$_0$ \& G$_2$ & $\pm 6$ & $^{+1}_{-11}$ & $\pm 0.4$ 
 & $^{+2}_{-1}$ & $^{+1}_{-6}$ & $^{+7}_{-6}$ \\ \hline 
\hline
Total & $^{+218}_{-25}$ & $^{+31}_{-129}$ & $^{+23.5}_{-5.2}$ 
& $^{+12}_{-73}$ & $^{+28}_{-192}$ & $^{+379}_{-68}$  \\
\hline  \hline 
\end{tabular}
\end{center}
\end{table}

The resulting systematic errors are summarized in Table~\ref{tab:syser}.
Total systematic errors are calculated by adding the individual errors in 
quadrature.
We obtain the mass, total width and 
$\Gamma_{\gamma \gamma} {\cal B}(\pi^0 \pi^0)$
of the $f_4(2050)$ to be 
$ 1884~^{+14}_{-13}~^{+218}_{-25}~\MeV/c^2$,
$453 \pm 20 ~^{+31}_{-129}~\MeV$ and 
$7.7~^{+1.2}_{-1.1}~^{+23.5}_{-5.2}~\eV$, 
where the first errors are statistical and the second systematic.
The errors are dominated by systematics, and mostly come from
uncertainties due to the unfolding procedure and
background parametrization, and possible biases
in the $\cos \theta^*$ distribution.

From the measured branching fraction to $\pi \pi$ (Table~\ref{tab:param}), 
the two-photon width of the $f_4(2050)$ is obtained to be 
$136~^{+24}_{-22}~^{+415}_{-91}~\eV$.
Given the large systematic error, we cannot conclude that
the two-photon width of the $f_4(2050)$ is nonzero.
However our data clearly require a G-wave component (see Fig.~\ref{fig:fig6}),
and the unacceptable fit without the $f_4(2050)$ 
(Table~\ref{tab:fit1})
strongly supports a finite two-photon coupling.
In the past, TASSO and JADE have set 95\% confidence upper limits 
for the $f_4(2050)$ to be 
$\Gamma_{\gamma \gamma} {\cal B}(K K) < 0.29~\keV$~\cite{tasso} and
$\Gamma_{\gamma \gamma} {\cal B}(\pi \pi) < 1.1~\keV$~\cite{jade}, 
respectively.
These can be translated into upper limits for the two-photon widths of
43 and 6.5~keV, respectively.
The power of such a large statistics (3 orders of magnitude more)
of our experiment is evident.
The nominal fit brings quite unexpected results: it requires
a "flip" of the $f_4(2050)$ and ``$f_2(1950)$'' positions with the mass
of the former becoming $1885^{+14}_{-13}$ MeV or 153 MeV lower than the
optimal mass of the ``$f_2(1950)$''. 
In addition, the fit requires
both states to be much broader than before, 440-450 MeV or about 2 times
larger than their PDG values. 
Obviously, the interference of the D$_2$ and G$_2$
amplitudes with each other and with the underlying continuum
demands a more sophisticated description probably involving more than one
resonance in each wave. 
Such a full amplitude analysis is beyond the scope
of this work. 
On the other hand, results of all the fits provide
unambiguous evidence for a nonzero two-photon coupling of the G wave.

\section{Analysis of the higher-energy region}
\label{sec:highe}
In general, we expect that at high energies and large scattering angles,
leading term QCD calculations give reasonable predictions for hard
exclusive processes such as $\gamma\gamma \to M_1M_2$, where $ M_{1(2)}$ are
mesons. 
However, at what energies these terms begin dominating depends
on the hadrons involved. 
In addition, even at the highest energies
the differential cross sections depends on the shape of the  $M_{1(2)}$ 
wave functions. 
For charged meson pairs such as $\pi^+\pi^-$, $K^+K^-$, the differential 
cross sections is only slightly sensitive
to the shape of the wave functions and the numerically largest term
in the differential cross sections is proportional to 
$\sin^{-4}{\theta^*}$~\cite{bl,bc,chern}.
However for neutral meson pairs this term is absent; the cross section
$d \sigma /d |\cos{\theta^*}|$ is much 
smaller and  much more sensitive to the shape of the meson wave 
functions~\cite{bl,bc,chern}.

In contrast, the main idea of the handbag model~\cite{handbag} is that 
the terms that are asymptotically power corrections, give the numerically 
largest contributions even at currently available energies.
The universal prediction of the handbag model is that the ratios
$d \sigma (M^0 \bar{M^0}) /d \sigma (M^+M^-)$ are constant, i.e. 
the energy and angular dependences are the same for charged and 
neutral mesons. 
In particular, 
$d\sigma (\pi^0 \pi^0) / d \sigma (\pi^+ \pi^-) = 0.5$~\cite{handbag}
while it varies from $\approx 0.07$ at $\cos{\theta^*}=0$ to $\approx 0.04$
at $|\cos \theta^*|=0.6$ in ~Ref.\cite{bc}.

\subsection{Angular dependence}
\begin{figure}
\centering
\includegraphics[width=8cm]{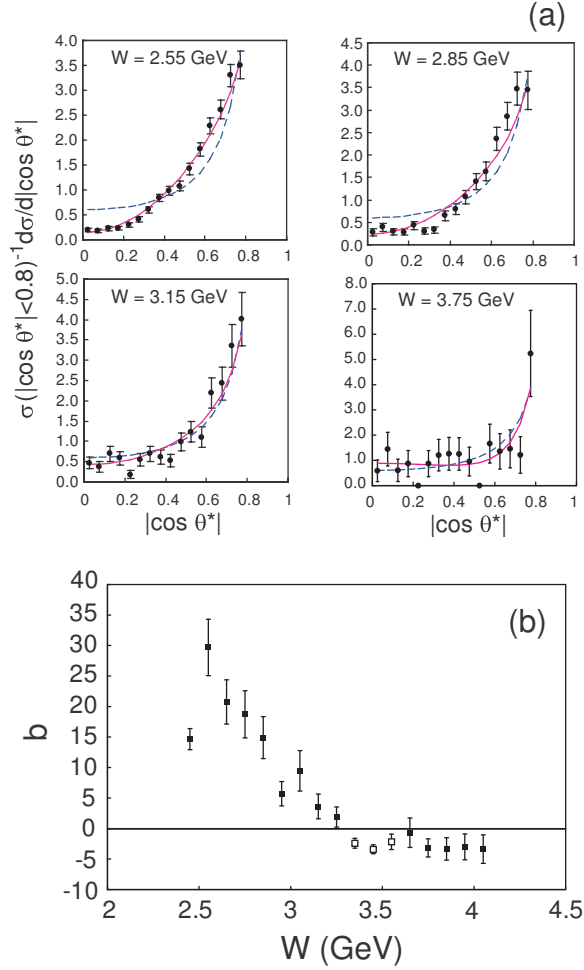}
\centering
\caption{(a) The fits of the angular dependence of the normalized differential 
cross sections (see text) at four selected $W$ points.
For the dashed curves the coefficient $b$ (see the fit formula in the text) 
is fixed to 0. 
The solid curves show the fits with $b$ floating. 
(b) The energy dependence of the parameter $b$ giving the best fits. 
Here, the charmonium contributions are not subtracted, and
the data in the $\chi_{c0}$ and $\chi_{c2}$ charmonium 
regions are plotted with open squares.}
\label{fig:fig8}
\end{figure}

We compare the angular dependence of the differential cross sections 
in the range $|\cos \theta^*|<0.8$ for
$W > 2.4~\GeV$ with the function $\sin^{-4} \theta^*$.
We also try a fit with an additional $\cos^2 \theta^*$ term, 
to quantify a possible deviation from the $\sin^{-4} \theta^*$ behavior.
We choose this function because it gives relatively good fits 
over a wide range in $W$.
Thus the fit function is parametrized as:
\begin{equation}
d\sigma/d|\cos \theta^*| = a(\sin^{-4} \theta^* + b\cos^2 \theta^*) .
\label{eqn:angul}
\end{equation}
We fit using a binned maximum likelihood method and 16 bins
in  the range $|\cos \theta^*|<0.8$.  
We know that the effect of charmonia is large in the region 
$3.3~\GeV < W < 3.6~\GeV$, but we cannot separate it
in the angular dependence because we cannot assume here any functional 
shapes for the noncharmonium component. 
The results of the fit for $b$ are shown in Fig.~\ref{fig:fig8},
as well as the fit to the angular distributions in
the four selected $W$ regions, 
where the differential cross sections, the vertical axis of this figure, 
are normalized to the total
cross section $\sigma(|\cos \theta^*|<0.8)$ in each $W$ region, i.e. 
the area under the curve is 1.
The parameter $b$ is close to zero above $W > 3.1~\GeV$
compared to $b \sim 10$, when the contribution of the
$b \cos^2 \theta^*$ term in the total cross section, 
$\sigma(|\cos \theta^*|<0.8)$,
is comparable to the contribution of the 
$\sin^{-4} \theta^*$ term.
The $b$ parameter becomes nearly constant and then systematically negative 
above the charmonium region.
The change in the $b$ parameter, which approaches a constant value near zero,
occurs at a $W$ value close to that observed in the charged-pion 
case~\cite{nkzw}.

\subsection{Yields of $\chi_{cJ}$ charmonia}
\begin{figure}
\centering
\includegraphics[width=10cm]{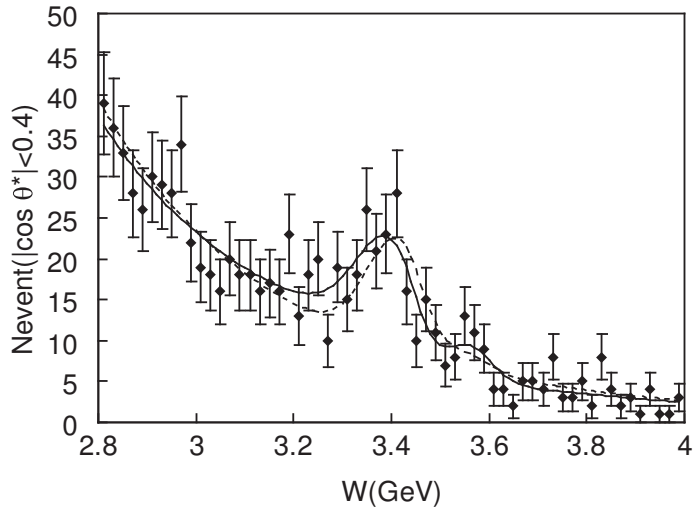}
\centering
\caption{The $W$ distribution
of the candidate events with $|\cos \theta^*|<0.4$  
near the charmonium region.
The solid and dashed curves show the fits described in the text with and
without interference with the $\chi_{c0}$.}
\label{fig:fig9}
\end{figure}

The structures seen in the yield distribution 
for $3.3~\GeV < W < 3.6~\GeV$ and $|\cos \theta^*| < 0.4$
(Fig.~\ref{fig:fig9}) are from charmonium production, 
$\gamma \gamma \to \chi_{c0}$, $\chi_{c2} 
\to \pi^0\pi^0$. 
Similar production of the two charmonium states
is observed in the $\pi^+\pi^-$, $K^+K^-$ and $K^0_SK^0_S$ final 
states~\cite{nkzw,wtchen}.
 
We fit the distribution to contributions from the $\chi_{c0}$,
$\chi_{c2}$ and a smooth continuum component using the following function: 
\begin{equation}
Y(W) = |\sqrt{\alpha kW^{-\beta}}+e^{i\phi}\sqrt{N_{\chi_{c0}}}
{\rm BW}_{\chi_{c0}}(W)|^2 + 
N_{\chi_{c2}}|{\rm BW}_{\chi_{c2}}(W)|^2 + \alpha (1-k)W^{-\beta},
\label{eqn:chicj}
\end{equation}
in the $W$ region between 2.8 and 4.0~GeV,
where ${\rm BW}_{\chi_{cJ}}(W)$ is a Breit-Wigner function for the charmonium
amplitude, which is proportional to 
$1/(W^2-M_{\chi_{cJ}}^2-iM_{\chi_{cJ}} \Gamma_{\chi_{cJ}})$
and is normalized as $\int |{\rm BW}_{\chi_{cJ}}(W)|^2dW=1$.
The masses and widths, $M$ and $\Gamma$, of the charmonium states
are fixed to the PDG world averages~\cite{pdg}.
The component $\alpha W^{-\beta}$ corresponds to the contribution from 
the continuum,
with a fraction $k$ that interferes with the $\chi_{c0}$ amplitude 
with a relative phase angle, $\phi$.  
It is impossible to determine the interference parameters for the
 $\chi_{c2}$ because of its much smaller intrinsic width compared 
to experimental resolution. 
We fit the $\chi_{c2}$ yield ($N_{\chi c2}$) with a formula where
no interference term is included, and later we estimate the maximum
effects from the interference term when determining the two-photon
decay width of $\chi_{c2}$.
We use data only in the range $|\cos \theta^*|<0.4$ where the charmonium 
contribution is dominant. 
Smearing effects 
due to a finite mass resolution 
are taken into account in the fit, using the same function
as used for the unfolding.

A binned maximum likelihood method is applied. 
We examined two cases with and without the interference. 
Reasonably good fits are obtained for both cases.
The fit results are summarized in Table~\ref{tab:charm1}. 
In the table, ${\cal L}$ is the likelihood value
and $ndf$ is the number of degrees of freedom.
The normalization $N_{\chi c0}$ in Eq.(8) is proportional to 
the square of the resonance amplitude.
The yields from the fits are translated into products 
of the two-photon
decay width and the branching fraction, 
$\Gamma_{\gamma \gamma}(\chi_{cJ}){\cal B}(\chi_{cJ} \to \pi^0\pi^0)$, 
which are listed in Table~\ref{tab:charm2}. 
The systematic errors are taken from the changes in the 
central values of the fitted yields when the absolute energy
scale is varied by $\pm 10$~MeV for the $W$ measurement, the
invariant-mass resolution is varied by $\pm 10$\% for the corresponding
Gaussian widths, and the fitting range is
narrowed to the range 2.96 - 3.84~GeV, and when the efficiencies
are varied by their uncertainties.
The changes in the goodness of fit ($-2\ln{\cal L}$) for 
the first two  variations are found to be small, less than 1.7, for
the interference case.

The $\chi_{c0}$ is observed with a statistical significance of $7.6\sigma$
($7.3\sigma$) when we take (do not take) interference into account.
The statistical significance for the $\chi_{c2}$ is $2.6\sigma$ when
we take interference of the $\chi_{c0}$ into account, but it is
only $1.3\sigma$ when we do not take into account interference. 
This is because interference makes the line shape
of the $\chi_{c0}$ highly asymmetric with a short tail and destructive 
interference on the high-energy side.
The solid and dashed curves in Fig.~\ref{fig:fig9}
show the fits for the two cases
(with and without $\chi_{c0}$ interference).

The results for  $\Gamma_{\gamma\gamma}{\cal B}(\chi_{cJ})$ in the
$\pi^0\pi^0$ final state can be compared to the only direct measurement
of this quantity in the $\pi^+\pi^-$ decay mode from Belle, 
$15.1 \pm 2.1 \pm 2.3$ eV and $0.76 \pm 0.14 \pm 0.11$ eV for the 
$\chi_{c0}$ and $\chi_{c2}$, respectively~\cite{nkzw}. 
Although the effects of interference were neglected in the $\pi^+\pi^-$
 measurements,
the results are consistent with the ratio expected from isospin invariance,
${\cal B}(\chi_{cJ} \to \pi^0\pi^0)/{\cal B}(\chi_{cJ} \to \pi^+\pi^-)=1:2$. 
Our results for $\Gamma_{\gamma\gamma}{\cal B}(\chi_{cJ})$ for the 
$\chi_{c0(2)}$ agree within errors with the indirect determination
of these quantities using the corresponding world averages~\cite{pdg}
or recent measurements of 
${\cal B}(\chi_{c0(2)} \to \pi^0\pi^0)$~\cite{cleobr1}
as well as  ${\cal B}(\chi_{cJ} \to \gamma\gamma)$ and 
$\Gamma_{\gamma\gamma}(\chi_{cJ})$~\cite{cleobr2} by 
the CLEO collaboration.

\small
\begin{center}
\begin{table}
\caption{Results of the fits (see text) to obtain the charmonium
contributions with and without interference effects. 
Errors are statistical only. Logarithmic likelihood
($\ln {\cal L}$) values are only meaningful when comparing two or more
fits.}
\label{tab:charm1}
\begin{tabular}{c|ccccc}
\hline
Interference & $N_{\chi_{c0}}$ & $k$ & $\phi$ &
$N_{\chi  c2}$ & $-2\ln{\cal L}/ndf$ \\
\hline
Without & $100 \pm 16$ & $\cdots$ & $\cdots$ & $13^{+11}_{-10}$ & $52.4/56$\\
With & $103^{+60}_{-42}$ & $0.82^{+0.18}_{-0.48}$ & $(1.1 \pm 0.3)\pi$ 
& $34 \pm 13$ & $44.2/54$\\
\hline
\end{tabular}
\end{table}
\end{center} 
\ \\
\begin{center}
\small
\begin{table}
\caption{Products of the two-photon decay width
and the branching fraction for the two charmonia.
Here,  $\Gamma_{\gamma \gamma}{\cal B}(\chi_{cJ})$
means $\Gamma_{\gamma \gamma}(\chi_{cJ}){\cal B}(\chi_{cJ} \to \pi^0\pi^0)$.
The first, second and third errors (when given) are statistical, systematic
and from the maximal uncertainties of the relative phase in $\chi_{c2}$
production.}
\label{tab:charm2}
\begin{tabular}{c|cc}
\hline
Interference & $\Gamma_{\gamma \gamma}{\cal B}(\chi_{c0})$ (eV)& 
$\Gamma_{\gamma \gamma}{\cal B}(\chi_{c2})$ (eV)\\
\hline
Without & $9.7 \pm 1.5 \pm 1.2$ & $0.18^{+0.15}_{-0.14} \pm 0.08$ \\
With & $9.9^{+5.8}_{-4.0} \pm 1.6$ & $0.48 \pm 0.18 \pm 0.07 \pm 0.14$ \\
\hline
\end{tabular}
\end{table}
\end{center} 
  
\normalsize

\subsection{Subtraction of the charmonium contributions}
We subtract the charmonium contributions from nearby bins
of the charmonium ($\chi_{cJ}$) region, $3.3 - 3.6~\GeV$,
in order to obtain a pure differential cross section
from the continuum component.
We use the fit result with interference obtained in the previous subsection. 

The estimated charmonium yield that includes the contribution from the
interference term is converted to a differential cross section contribution 
in each angular bin of $|\cos \theta^*|<0.8$ 
by assuming a flat distribution 
for the $\chi_{c0}$ component and a distribution $\sim \sin^4 \theta^*$ 
for the $\chi_{c2}$ component~\cite{wtchen}.
This assumption is only a model. 
In fact, we do not know the angular distribution
of the interference term; the charmonium amplitudes 
can interfere with the continuum 
components with different $J$'s of unknown sizes.

 For the $W=3.25$~GeV bin, the fit result indicates that there is
a non-negligible effect from the $\chi_{c0}$ when we assume interference,
and thus we make a correction for charmonium subtraction.  
The contribution of the charmonium components in the original differential 
cross sections is 18\% at $|\cos \theta^*|<0.6$. 
For  $W=3.3 - 3.6$~GeV,  we apply a subtraction  
for the angular bins  $0.4<|\cos \theta^*| <0.8$ after extrapolating 
the charmonium yield determined in the range $|\cos \theta^*| <0.4$.

The differential cross section thus obtained for the continuum is integrated 
over the range $|\cos \theta^*|<0.6$.
We convert $\sigma(0.4<|\cos \theta^*| <0.8)$ to  
$\sigma(|\cos \theta^*| <0.4)$ for $W=3.2 - 3.6$~GeV, 
by assuming that the angular dependence of the
differential cross section is $\sim \sin^{-4} \theta^*$. 
The results are plotted in Fig.~\ref{fig:fig10}(a), where
the cross section for $\gamma\gamma \to \pi^+\pi^-$ from Ref.~\cite{nkzw} 
is also shown.

\subsection{$W$ dependence and ratio of cross sections of 
$\pi^0 \pi^0$ to $\pi^+\pi^-$}
We fit the differential cross sections integrated over the polar angle, 
$\sigma(|\cos \theta^*|<0.6)$, 
to a power law in the c.m. energy, $W^{-n}$, for the energy region 
$3.1~\GeV <W< 4.1~\GeV$, in which the angular
dependence of the differential cross section does not show any large changes. 
In the fit, we do not use the data in the
charmonium region ($W=3.3 - 3.6$~GeV), where we cannot determine 
the cross section
of the continuum component in a model-independent manner.

The result of the power-law fit is 
summarized in Table~\ref{tab:val_n} and compared to that for 
other processes.
The systematic error is dominated by the uncertainty of the charmonium
contribution in the range $3.1~\GeV < W < 3.3~\GeV$.
This $n$ value is compatible with the results for the $\pi^+\pi^-$ 
and $K^+K^-$ processes~\cite{nkzw},
but significantly smaller than that
in the  $K^0_S K^0_S$ case~\cite{wtchen}. 
\begin{center}
\begin{table}
\caption{The value $n$ in $\sigma_{\rm tot} \propto W^{-n}$ in
various reactions fitted in the $W$ and $|\cos \theta^*|$ ranges indicated.}
\label{tab:val_n}
\begin{tabular}{lcccc} \hline \hline
Process & $n$ & $W$ range (GeV) & $|\cos \theta^*|$ range & Reference
\\ \hline
$\pi^0\pi^0$ & $6.9 \pm 0.6 \pm 0.7$ & ~~~~3.1 -- 4.1 (exclude 3.3 -- 3.6) 
& $<0.6$ & This exp't \\
$\pi^+\pi^-$ & $7.9 \pm 0.4 \pm 1.5$ & 3.0 -- 4.1 & $<0.6$ & \cite{nkzw} \\
$K^+K^-$  & $7.3 \pm 0.3 \pm 1.5$ & 3.0 -- 4.1 & $<0.6$ & \cite{nkzw} \\
$K^0_S K^0_S$  & $10.5 \pm 0.6 \pm 0.5$ & 2.4 -- 4.0 (exclude 3.3 -- 3.6) 
& $<0.6$ & \cite{wtchen} \\
\hline
$\pi^0\pi^0$ & $8.0 \pm 0.5 \pm 0.4$ & 3.1 -- 4.1 (exclude 3.3 -- 3.6) 
& $<0.8$ & This exp't \\
\hline\hline
\end{tabular}
\end{table}
\end{center}

\begin{figure}
\centering
\includegraphics[width=8cm]{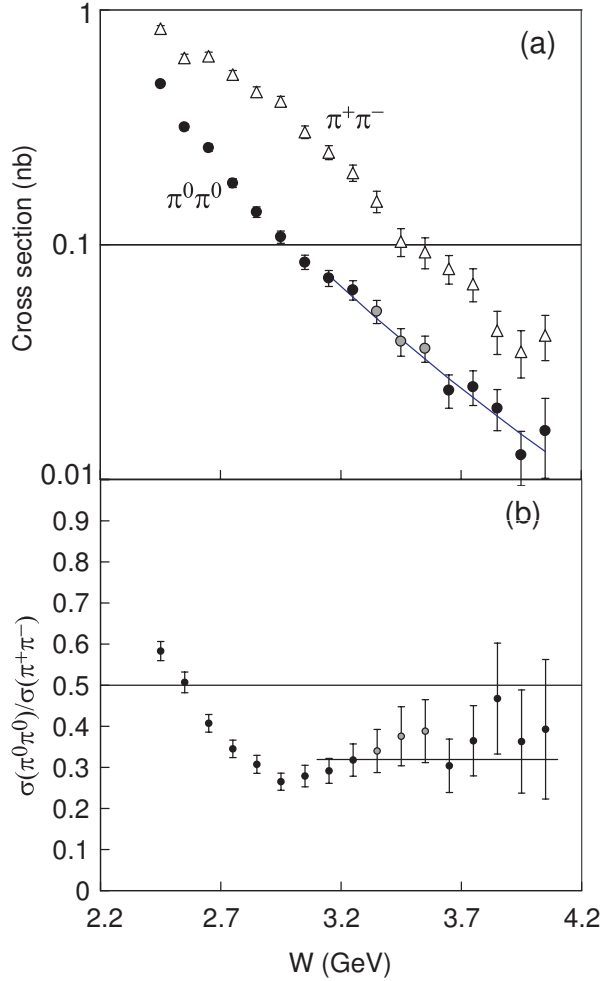}
\centering
\caption{(a) The cross sections for the
$\gamma\gamma \to \pi^0\pi^0$ (solid circles) and
$\gamma \gamma \to \pi^+\pi^-$ (triangles, \cite{nkzw})
for $|\cos \theta^*|<0.6$.
The curve is a fit
to the cross section for $\gamma\gamma \to \pi^0\pi^0$ 
with a $\sim W^{-n}$ functional shape.
(b) Ratio of the cross section for the $\pi^0\pi^0$ process
to the $\pi^+\pi^-$ process. 
The error bars are statistical only. 
The shorter horizontal line is the average for $3.1~\GeV < W < 4.1~\GeV$.
The horizontal line (0.5) is an expectation
from isospin invariance for a pure $I=0$ component.
In (a) and (b), the estimated charmonium contributions 
are subtracted in both $\pi^+\pi^-$ and $\pi^0\pi^0$ measurements.
The results in the $W$ region 3.3 - 3.6~GeV (plotted 
with gray circles) are not used for the fits.}
\label{fig:fig10}
\end{figure}

The fit for  $3.1~\GeV < W < 4.1~\GeV$ is shown in Fig.~\ref{fig:fig10}(a),
which also shows the cross section of $\pi^0\pi^0$ from the Belle
measurement~\cite{nkzw}.
In Fig.~\ref{fig:fig10}(b) we show the ratio of the cross sections of
$\pi^0 \pi^0$ to $\pi^+\pi^-$. 
This ratio is rapidly falling at low energies, but its behavior changes
above 3.1~GeV, where the two processes have similar $W^{-n}$ dependence,
which results in the almost constant ratio.
The average of the ratio in this energy region is 
$0.32 \pm 0.03 \pm 0.05$, where the data in the 
3.3 - 3.6~GeV region is not used when calculating this average.
This ratio is significantly larger than the prediction of the leading-order
QCD calculations~\cite{bl,bc,chern}
and is somewhat smaller than the value of 0.5, 
which is suggested by isospin invariance~\cite{handbag}.

\section{Summary and Conclusion}
\label{sec:concl}
We have measured the process $\gamma \gamma \to \pi^0\pi^0$ using
a high-statistics data sample 
from $e^+e^-$ collisions corresponding to an integrated luminosity
of 223~fb$^{-1}$ collected
with the Belle detector at the KEKB accelerator. 
We derive results for the differential cross sections
in the center-of-mass energy and polar angle ranges, 
$0.6~\GeV < W < 4.1~\GeV$ and $|\cos \theta^*|<0.8$. 

Differential cross sections are fitted in the energy region
$1.7~\GeV < W < 2.5~\GeV$ in a model where the partial waves consist of
resonances and smooth backgrounds. 
Various fits are performed that provide
unambiguous evidence for a nonzero two-photon coupling of the G wave.
Helicity-2 production (G$_2$) is preferred compared to the
helicity-0 (G$_0$) one.

We observe production of the charmonium state $\chi_{c0}$ and obtain the
product of its two-photon decay width and the branching fraction to 
$\pi^0\pi^0$.
The angular distribution of the differential cross section is largely
energy dependent, and approaches $\sim \sin^{-4} \theta^*$ 
above $W=3.1$~GeV.  This observation and the energy
dependence of the cross section above this energy, which is well fitted by
$W^{-n}$, $n=6.9 \pm 0.6 \pm 0.7$, 
are compatible with those measured in the $\pi^+\pi^-$ channel.
We obtain the cross section ratio,
$\sigma(\pi^0\pi^0)/\sigma(\pi^+\pi^-)$, to be $0.32 \pm 0.03 \pm 0.05$ 
on average in the 3.1-4.1~GeV region. 
This ratio is significantly larger than the prediction of the leading-order
QCD calculation.

\section*{Acknowledgment}
We are grateful to V. Chernyak for useful discussions.
We thank the KEKB group for the excellent operation of the
accelerator, the KEK cryogenics group for the efficient
operation of the solenoid, and the KEK computer group and
the National Institute of Informatics for valuable computing
and SINET3 network support.  We acknowledge support from
the Ministry of Education, Culture, Sports, Science, and
Technology (MEXT) of Japan, the Japan Society for the 
Promotion of Science (JSPS), and the Tau-Lepton Physics 
Research Center of Nagoya University; 
the Australian Research Council and the Australian 
Department of Industry, Innovation, Science and Research;
the National Natural Science Foundation of China under
contract No.~10575109, 10775142, 10875115 and 10825524; 
the Department of Science and Technology of India; 
the BK21 program of the Ministry of Education of Korea, 
the CHEP src program and Basic Research program (grant 
No. R01-2008-000-10477-0) of the 
Korea Science and Engineering Foundation;
the Polish Ministry of Science and Higher Education;
the Ministry of Education and Science of the Russian
Federation and the Russian Federal Agency for Atomic Energy;
the Slovenian Research Agency;  the Swiss
National Science Foundation; the National Science Council
and the Ministry of Education of Taiwan; and the U.S.\
Department of Energy.
This work is supported by a Grant-in-Aid from MEXT for 
Science Research in a Priority Area ("New Development of 
Flavor Physics"), and from JSPS for Creative Scientific 
Research ("Evolution of Tau-lepton Physics").

\end{document}